\documentclass[prl,nofootinbib,twocolumn,elsart12]{revtex4}
\usepackage{graphicx}
\usepackage{bm}
\usepackage{float}
\usepackage{subfigure}
\usepackage{amsmath}
\usepackage{amsfonts}
\usepackage{epsfig}
\usepackage{hyperref}
\usepackage{mathrsfs}

\def\half{{\mathchoice{{\textstyle{1\over 2}}}{1\over 2}{1\over 2}{1
\over 2}}}

\def\Phidag{\Phi^\dagger}

\def\({\left(}
\def\){\right)}
\def\[{\left[}
\def\]{\right]}

\def\CC{{\rm I\!\!\! C}}

\def\bed{\begin{description}}
\def\eed{\end{description}}

\def\ba{\begin{array}}
\def\ea{\end{array}}

\def\half{\frac{1}{2}\,}

\def\u1{$U(1)$}
\def\suu1{$SU(2)\times U(1)$}

\def\nn{\nonumber}

\def\({\left(}
\def\){\right)}
\def\[{\left[}
\def\]{\right]}
\def\Phidag{\Phi^\dagger}
\def\CC{{\rm I\!\!\! C}}

\begin{document}
\newcommand{\newc}{\newcommand}
\newc{\be}{\begin{equation}}
\newc{\ee}{\end{equation}}
\newc{\bear}{\begin{eqnarray}}
\newc{\eear}{\end{eqnarray}}
\newc{\bea}{\begin{eqnarray*}}
\newc{\eea}{\end{eqnarray*}}
\newc{\D}{\partial}
\newc{\ie}{{\it i.e.} }
\newc{\eg}{{\it e.g.} }
\newc{\etc}{{\it etc.} }
{\newc{\etal}{{\it et al.}}
\newc{\lcdm}{$\Lambda$CDM}
\newc{\ra}{\rightarrow}
\newc{\lra}{\leftrightarrow}
\newc{\lsim}{\buildrel{<}\over{\sim}}
\newc{\gsim}{\buildrel{>}\over{\sim}}
\newcommand{\fs}{{\rm{\it f\sigma_8}}}
\newcommand{\mincir}{\raise
-3.truept\hbox{\rlap{\hbox{$\sim$}}\raise4.truept\hbox{$<$}\ }}
\newcommand{\magcir}{\raise
-3.truept\hbox{\rlap{\hbox{$\sim$}}\raise4.truept\hbox{$>$}\ }}

\title{Dilatonic Topological Defects in 3+1 Dimensions and their Embeddings}

\author{Nikos Platis}\email{n\_platis@yahoo.com}

\author{Ioannis Antoniou}\email{ianton@cc.uoi.gr}

\author{Leandros Perivolaropoulos}\email{leandros@uoi.gr}
\affiliation{Department of Physics, University of Ioannina, Greece}

\date{\today}

\begin{abstract}
We consider Lagrangians in 3+1 dimensions admitting topological defects where there is an additional coupling between the defect scalar field $\Phi$ and the gauge field kinetic term (eg $B(\vert \Phi \vert^2) F_{\mu \nu}F^{\mu \nu}$). Such a {\it dilatonic} coupling in the context of a static defect, induces a spatially dependent effective gauge charge and effective mass for the scalar field which leads to modified properties of the defect core. In particular, the scale of the core gets modified while the stability properties of the corresponding embedded defects are also affected. These modifications are illustrated for gauged (Nielsen-Olesen) vortices and for gauged ('t Hooft-Polyakov) monopoles. The corresponding dilatonic global defects are also studied in the presence of an external gauge field.
\end{abstract}
\maketitle

\section{Introduction}

The spacetime variation of fundamental constants\cite{Damour:2009zy,Uzan:2010pm,Flambaum:2007my} like the gravitational constant\cite{brans1} or the charges of gauge field interactions\cite{Teller} is usually implemented at the Lagrangian level by promoting these constants to scalar fields\cite{brans1,Teller,Bekenstein1} whose dynamics is determined by  potential and kinetic terms properly chosen to make the allowed variations consistent with current experiments and cosmological observations. For example in order to allow spacetime variation of the gravitational constant $G$ and the fine structure constant $\alpha= e^2/{\hbar c}$ we may consider the replacement of the Einstein-Maxwell action
\be
S=\int d^4 x \sqrt{-g} \left(\frac{c^4}{16\pi G_0} R - \frac{1}{4\alpha_0} F_{\mu \nu} F^{\mu \nu} + {\cal L}_m \right)
\label{einst}
\ee
by a generalization
\begin{widetext}
\be
S=\int d^4 x \sqrt{-g} \left(\phi R -\omega_\phi \frac{\phi_{,\mu}\phi^{,\mu}}{\phi} - V_\phi(\phi) - e^{-2\psi} \frac{1}{4\alpha_0} F_{\mu \nu} F^{\mu \nu} -\frac{\omega_\psi}{2} \psi_{,\mu}\psi^{,\mu} - V_\psi(\psi) \right)
\label{bdbsbm}
\ee
\end{widetext}
inspired from the Brans-Dicke (BD) theory\cite{brans1} (gravitational part) and the  Bekenstein \cite{Bekenstein1} Sandvik, Barrow and Magueijo \cite{Sandvik1} BSBM (electromagnetic part) actions.
For free fields, the potentials take the forms: $V(\phi)=\frac{1}{2} m_\phi^2 \phi^2$, $V(\psi)=\frac{1}{2} m_\psi^2 \psi^2$.
In this action the gravitational constant $G_0$ is replaced by the dynamical BD field $\phi$ as $\phi=\frac{16\pi G}{c^4}$ and the fine structure constant is replaced by the dynamical BSBM field $\psi$ as $\alpha=\alpha_0 e^{2\psi}$. Laboratory experiments and astrophysical/cosmological observations impose limits on the allowed spacetime variations of $G$ \cite{Will:2005va} and $\alpha$ \cite{Chiba:2011bz}. These limits can be translated into constraints on the parameters $\omega_\phi$, $\omega_{\psi}$ and on the masses of the corresponding scalar fields. In the limit of infinite values of these parameters, the dynamics of the scalar fields freeze and the dynamics of the action (\ref{bdbsbm}) reduces to the Einstein-Maxwell action dynamics. The scalar fields $\phi$ and $\psi$ emerge naturally in the context of string theory as {\it dilatons} \cite{Piazza:2004df,Damour:2002nv}.

The BD parameter $\omega_\phi$ is dimensionless while the BSBM parameter $\omega_\psi$ has dimensions of energy squared $m^2\sim l^{-2}$ (in units where $h=c=1$). If the potentials are ignored ($m_\phi=m_{\psi}=0$) then the experimental/observational constraints on $\omega_\phi$, $\omega_\psi$ are   \cite{Will:2005va,Chiba:2011bz,Farajollahi:2012mw}
\bea
\omega_\phi &>& 4 \times 10^4 \\
(100 MeV)^2 &<& \omega_\psi < M_{Pl}^2
\eea
These constraints are based mainly on tests of the equivalence principle and fifth force search experiments as well as on solar system tests (for $\omega_\phi$).
When the field masses are nonzero, the above constraints are significantly relaxed \cite{Perivolaropoulos:2009ak}.

In a cosmological setup both fields $\phi$ and $\psi$ have been considered as possible dark energy candidates \cite{Copeland:2006wr,Farajollahi:2012mw,Chiba:2001er,Barrow:2013uza,Copeland:2003cv,Bertolami:2003qs,Lee:2004vm,Calabrese:2013lga}. In the context of the recent possible detection of temporal\cite{Murphyc,Webb,webb1,Webb2,webb3,webb4,webb5} and spatial \cite{King:2012id} variation of $\alpha$ on cosmological scales (the $\alpha$ dipole \cite{King:2012id,Mariano:2012wx}), the field $\psi$ has the potential to play a dual role: the role of inhomogeneous dark energy and the cause of $\alpha$ variation \cite{Copeland:2006wr,Farajollahi:2012mw,Chiba:2001er,Barrow:2013uza,Copeland:2003cv,Bertolami:2003qs,Lee:2004vm}. Cosmological models based on inhomogeneous dark energy are motivated by CMB and other cosmic anomalies \cite{Perivolaropoulos:2014lua} which may hint towards deviations from the cosmological principle on large cosmic scales \cite{Grande:2011hm,Perivolaropoulos:2014lua,Perivolaropoulos:2012mca}.

Negative pressure and large sound velocity would tend to wipe out any inhomogeneities of this scalar field on all scales. Topologically non-trivial field configurations however have the potential to sustain such field inhomogeneities on cosmological scales. Such configurations have been considered as a possible mechanism to sustain inhomogeneous dark energy (topological quintessence \cite{Grande:2011hm,BuenoSanchez:2011wr,Perivolaropoulos:2012mca}) possibly combined with correlated spatial variation of fine structure constant (extended topological quintessence \cite{Perivolaropoulos:2014lua,Mariano:2012ia}). The later possibility is amplified by the observational fact that a dipole fit of the dark energy distribution using Type Ia supernovae leads to a dipole whose direction is only about $10^\circ$ away from the $\alpha$ dipole direction \cite{Mariano:2012wx}. Therefore, topological defects emerging due to topologically non-trivial configurations of the field $\psi$ (dilatonic defects) have the potential to play an interesting role in cosmology \cite{Olive:2010vh,MenezesdaSilva:2007vj,Menezes:2005tp,Perivolaropoulos:2013uea,Olive:2012ck,Perivolaropoulos:2014lua}. It is therefore interesting to investigate their field configuration properties which emerge as generalizations of the corresponding ordinary defects where there is no coupling between the scalar field and the gauge field kinetic term. These properties may be summarized as follows:
\begin{itemize}
\item
The dilatonic coupling induces spatial variation of the gauge charge and a spatial variation of the effective mass of the scalar field. This can lead to modification of the scale of the gauged topological defect core.
\item
The stability of the gauged embedded defects \cite{Vachaspati:1991dz,Hindmarsh:1991jq,James:1992wb,James:1992zp,Barriola:1993fy,Achucarro:1999it} is significantly affected by the dilatonic coupling due to the spatial variation of the effective mass of the scalar field \cite{Perivolaropoulos:2013uea}.
\item
The dilatonic coupling can lead to the formation of a scalar field condensate in the core of embedded defects because it can induce a local instability which is confined in the core region where the gauge fields are excited.
\item
Global embedded defects are unstable without the dilatonic coupling. However, in the presence of a dilatonic coupling and an external gauge field, they can be locally stabilised in their core region.
\end{itemize}
The goal of the present study is to demonstrate the existence of dilatonic defect solutions and investigate in some detail the above properties in the case of dilatonic vortices and monopoles. This is an extension of a recent study by two of the authors that focused only on the case of the dilatonic semilocal (embedded) vortex \cite{Perivolaropoulos:2013uea}.

The structure of this paper is the following: In the next section we briefly review the Nielsen-Olesen U(1) gauged vortex and demonstrate the existence of its dilatonic generalization. The embedded dilatonic NO vortex is also defined and its stability is reviewed. In the limit of a trivial dilatonic coupling the embedded dilatonic gauged vortex reduces to the semilocal string. The embedded global dilatonic vortex is also defined and its stability is analyzed in the context of a localized external gauge (``magnetic'') field. In section 3 we define the dilatonic 'tHooft-Polyakov monopole and consider its embedding in a model with $O(4)$ symmetry. The embedded global dilatonic monopole is also defined and its stability is analyzed in the context of a localized external gauge (``magnetic'') field. Finally, in section 4 we conclude and discuss future extensions of the present analysis paying special attention to applications in the case of the dilatonic electroweak string.

\section{Dilatonic Vortices}

\subsection{Dilatonic Nielsen-Olesen Vortex}

We start this section with a brief review of the Nielsen-Olesen (NO) vortices without a dilatonic coupling. These are topologically stable string solutions in the {\it{Abelian-Higgs}} model \cite{NieOle73,Peri93}.
\\
The Lagrangian density of this model is of the form
\be
\mathscr{L}=- \frac{1}{4 e^2} F_{\mu \nu} F^{\mu \nu} + |D_{\mu} \Phi|^2 - V(\Phi) \label{abelianLagrangian}
\ee
where $D_{\mu} = \D_{\mu} - i A_{\mu}$ is the covariant derivative. Also $V(\Phi)=\frac{\lambda}{4}(\Phi^*\Phi - \eta^2)^2$.
The Nielsen-Olesen vortex ansatz is of the form
\bear
\Phi & = & f(r) e^{im \theta} \label{NO1} \\
A_{\mu} & = & A_{\theta}= \frac{u(r)}{r} \hat{\theta} \label{NO2}
\eear
where $m$ is the winding number of the complex scalar field $\Phi$.
\\
The vacuum manifold of the model is an $S^1$
\be
{\cal V}  = \{ \Phi \in \CC \  | \  \Phi^* \Phi - \eta^2  = 0 \}
\cong S^1 \ .
\label{vacman}
\ee
When the asymptotic value of the scalar field $\Phi$ wraps around that vacuum manifold, there is a  non-zero value of the winding number of the vortex. This forces the scalar field to acquire a zero value somewhere in the (xy) plane. The resulting vortices are {\it{topological}} and they are labelled by non-trivial elements of the first homotopy group: $\pi_1( {\cal V}) = \pi_1(S^1) \neq 1$.
\\
The energy momentum tensor
\be
T_{\mu \nu}=-g_{\mu \nu} \mathscr{L} +2 \frac{\D \mathscr{L}}{\D g^{\mu \nu}} \label{Tmn}
\ee
is easily obtained by using (\ref{abelianLagrangian}) in (\ref{Tmn}) with the NO ansatz (\ref{NO1}), (\ref{NO2}). Thus, the energy density is obtained as
\bear
\rho & = & f'^2 + \frac{f^2}{r^2}(m-u)^2 + \frac{u'^2}{2r^2} + \frac{\beta}{2}(f^2 - 1)^2
\eear
where the following rescaling was applied:
\bear
f \to \bar{f} & = & \eta f \label{rescale1} \\
r \to \bar{r} & = & \frac{r}{\eta e} \label{rescale3}
\eear
and $\beta\equiv  \frac{m_{\Phi}^2}{m_A^2}=\frac{\lambda}{2 e^2}$. Note that the scalar field mass is  $m_{\Phi}=\frac{\sqrt{\lambda}\eta}{\sqrt{2}}$ while the gauge field mass is $m_A=e \eta$. Thus, $\beta$ is the only free parameter of the theory after the rescaling.
Also the rescaled field equations obtained from ({\ref{abelianLagrangian}) are:
\bear
f'' + \frac{f'}{r} - \frac{f}{r^2}(m - u)^2 - \beta(f^2 -1)f & = & 0 \label{f_NO}\\
u'' - \frac{u'}{r} + 2 f^2 (m-u) & = & 0 \label{u_NO}
\eear
The NO boundary conditions to be imposed on (\ref{f_NO}) and (\ref{u_NO}) for winding $m$ are
$f(0)=u(0)=0, f(r\to \infty)=1$ and $u(r\to \infty)=m$. In this study we focus on vortices with unit winding number ($m=1$).
\\
In Fig. \ref{fig:NOsolutions(b=5)} we show the solution of the field equations (\ref{f_NO}) and (\ref{u_NO}) with the NO boundary conditions numerically for $\beta=5$

\begin{figure}[!ht]
\centering
\includegraphics[scale=0.35]{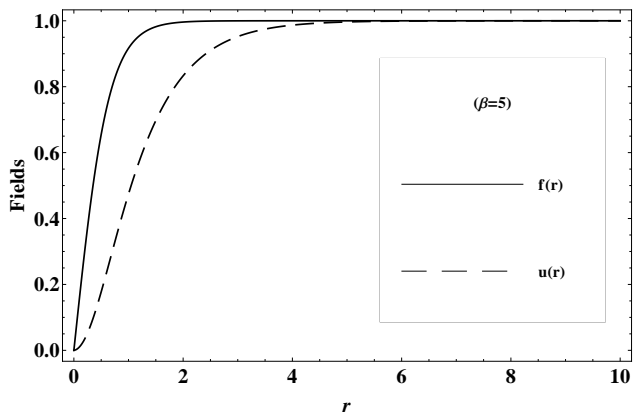}
\caption{The functions $f_{NO}$ and $u_{NO}$ for a string with unit winding $(m=1)$ and for $\beta=5$}
\label{fig:NOsolutions(b=5)}
\end{figure}

In the BPS limit ($\beta=1$) (following \cite{Hin93} and regrouping the terms of the total energy), the rescaled field eqs (\ref{f_NO})-(\ref{u_NO}) reduce to the following first order BPS equations
\be
f' + \frac{u-1}{r}f=0 \qquad\qquad u' + r(f^2 -1)= 0
\ee

We now add a dilatonic coupling of the form $B(|\Phi|^2)=1+q\frac{|\Phi|^2}{\eta^2}$ to the gauge kinetic term of the Abelian-Higgs model Lagrangian (\ref{abelianLagrangian}) thus generalizing the model to the {\it dilatonic Abelian-Higgs} model
\be
\mathscr{L} =  |D_{\mu} \Phi|^2 - \frac{B(\Phi)}{4 \alpha_0}F_{\mu \nu}F^{\mu \nu} - \frac{\lambda}{4}(\Phi^{*} \Phi - \eta^2)^2 \label{DilatonicLagrangian}
\ee
Thus, the fine structure constant $\alpha$ becomes dynamical with $\alpha = \alpha_0/B(\Phi)$. Notice that this coupling is not the same BSBM coupling of eq. (\ref{bdbsbm}) even though the two couplings are similar for small field values. We use this form to demonstrate the effect of the dilatonic term on the effective mass of the scalar field.
Using the same NO ansatz (\ref{NO1}), (\ref{NO2}) the rescaled energy density takes the form
\bear
\rho & = & f'^2 + \frac{f^2}{r^2}(1-u)^2 + \frac{1+qf^2}{2} \left(\frac{u'}{r} \right)^2 \nn \\
&& \:+ \frac{\beta}{2} (f^2 - 1)^2 \label{DilNOVorEnden}
\eear
and the field eqs are obtained as
\bear
&&f'' + \frac{f'}{r} - \frac{f}{r^2}(1 - u)^2 - \frac{q f}{2} \left(\frac{u'}{r} \right)^2- \nn \\
&&  \:\:\: - \beta (f^2 - 1) f  =  0  \\
&&u''- \frac{u'}{r} + \frac{2 f^2}{1 + q f^2}(1-u)  =  0
\eear
where the (\ref{rescale1}, \ref{rescale3}) rescaling was used.
\\
It is straightforward to show that for small $r$, the functions $f$ and $u$ behave as $f\sim r$ and $u\sim r^2$ respectively. As $r\to \infty$ they approach their asymptotic values exponentially with a width $w\sim \beta^{-1/2}$ for $f$ while for $u$ the width is independent of $\beta$ \cite{Peri93}.

In the presence of the dilatonic coupling we may define an effective rescaled mass squared (negative due to symmetry breaking) for the scalar field in the core region as
\bear
-\beta_{eff} & = & -\beta + \frac{q u'^2}{2r^2}  \label{effectivemass}
\eear
For $r<<1$, the term $\frac{q u'^2}{2r^2}$ is $O(q)$ while for $r>>1$ it vanishes. Thus, for $q>0$ ($q<0$),  $\beta_{eff}$ in the core region decreases (increases) leading to increased (decreased) core width. This is demonstrated in Fig. (\ref{fig:Effectivemassfields}) where the width of the dilatonic vortex increases as we increase the value of $q$.
\begin{figure}[!ht]
\centering
\includegraphics[scale=0.34]{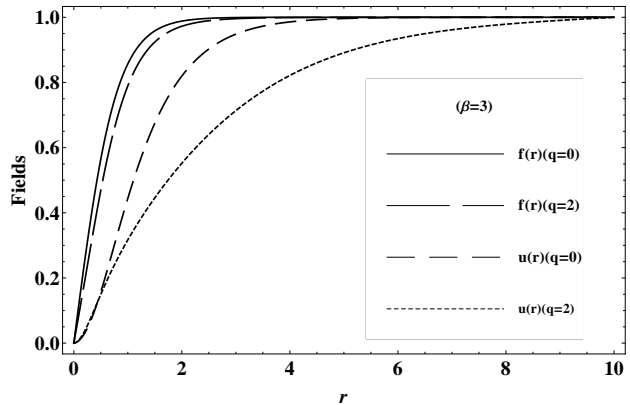}
\caption{As we increase $q$ the effective mass of the scalar field decreases, which leads to increased core width, as discussed above.}
\label{fig:Effectivemassfields}
\end{figure}
The rescaled effective mass squared $-\beta_{eff}$, shown in Fig. \ref{fig:Effectivemass}, significantly increases for $r<1$.

\begin{figure}[!ht]
\centering
\includegraphics[scale=0.365]{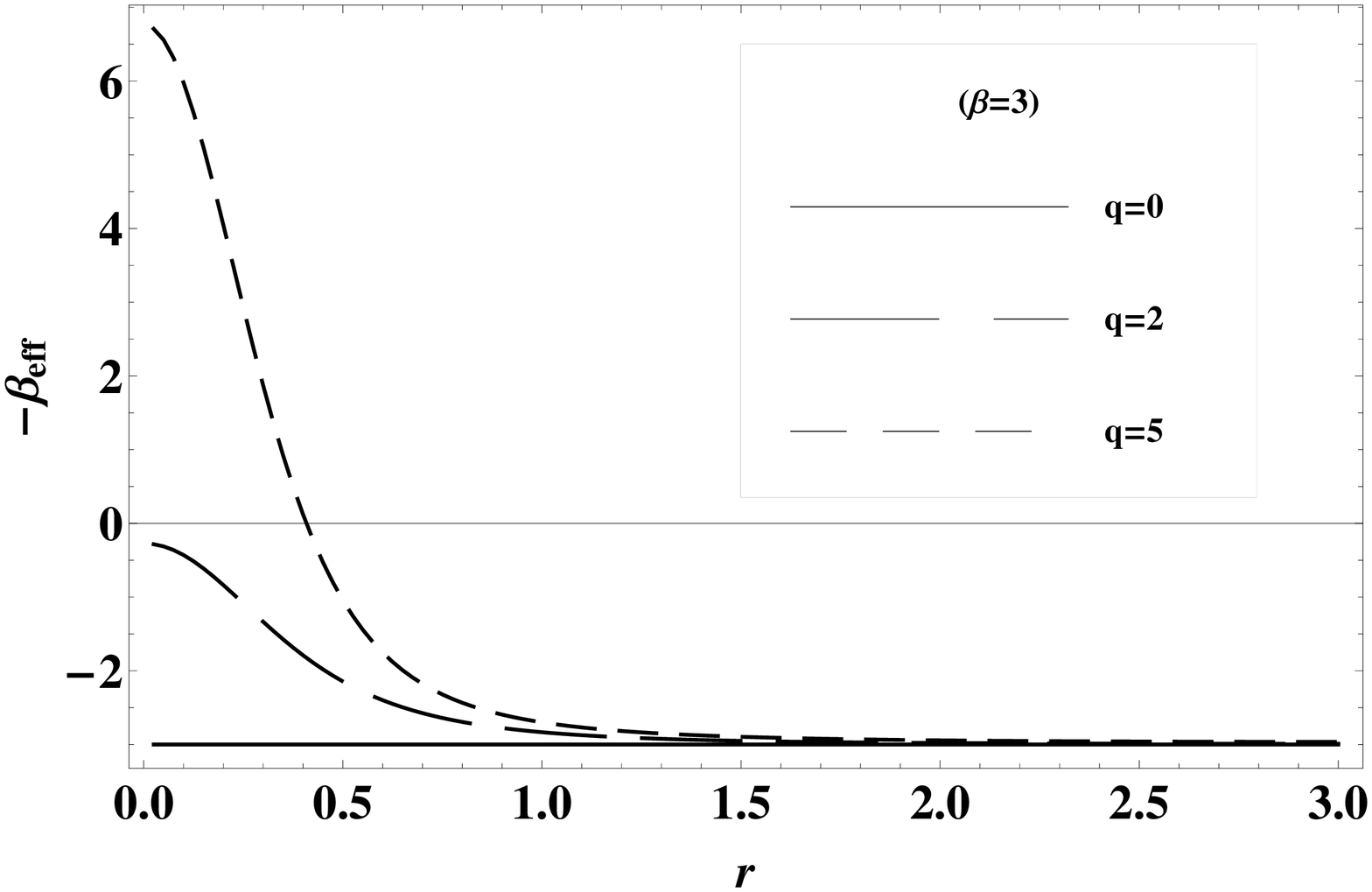}
\caption{The effective mass of the Higgs field $-\beta_{eff}$ changes locally around the core where $\frac{q u'^2}{2 r^2}$ is non-zero.}
\label{fig:Effectivemass}
\end{figure}

\subsection{Embedded Dilatonic Nielsen-Olesen Vortex}

The NO vortex can be embedded in various generalizations of the Abelian-Higgs model including the {\it{semilocal}} model \cite{VacAch91}. This is obtained if we promote the $U(1)_{gauge}$ symmetry to $SU(2)_{global}\times U(1)_{local}$. This is achieved by replacing the complex scalar field $\Phi$ by a complex $SU(2)$ doublet.
\\
The Langrangian density is of the form
\be
\mathscr{L} = (D_{\mu} \Phi)^\dag (D^{\mu}\Phi) - \frac{B(\Phi^{\dag}\Phi)}{4 \alpha_0}F_{\mu \nu}F^{\mu \nu} - \frac{\lambda}{4}(\Phi^{\dag} \Phi - \eta^2)^2
\ee
The embedded NO vortex ansatz is of the form
\be
\Phi= \left(
\begin{array}{c}
\Phi_1 \\
\Phi_2\\
\end{array} \right) =
\left(
\begin{array}{c}
0 \\
f e^{i \theta}\\
\end{array} \right) \label{doublet}
\ee
while for the gauge field remains unchanged (\ref{NO1}).
This time the vacuum manifold of the configuration is the 3-sphere
\be
{\cal V} \ = \
\{\Phi \in \CC^2 \ | \ \Phidag \Phi = \eta^2  \}
\
\cong
\ S^3 \ \ ,
\ee
Due to trivial topology of the vacuum in this case ($\pi_1 (S^3)=1$), the embedded NO vortex can only be dynamically stable with respect to small perturbations. The stability analysis of the embedded dilatonic NO vortex was presented in \cite{Perivolaropoulos:2013uea}. Here we briefly sketch the analysis for completeness. The range of stability for $q=0$ is $0<\beta<1$ \cite{Hin92,Hin93,AchKuiPerVac92,AchBorLid98}.

The only type of perturbation that is likely to lead to instability is of the form $\delta \Phi_{1}=g$. We focus on a real $g$ as any added phase would tend to increase the energy of the perturbations \cite{Achucarro:1999it}.
\\
The perturbed energy density is
\bear
\rho & = &  g'^2 + f'^2 + \frac{f^2}{r^2}(1-u)^2 + \frac{u^2 g^2}{r^2} \nn \\
&& \:+ \frac{1}{2}(1+q(f^2+g^2))\left(\frac{u'}{r} \right)^2 \nn \\
&& \:+ \frac{\beta}{2}(f^2 + g^2 -1)^2 \label{embdilNOenden}
\eear
and the field eqs are
\bear
&& u'' - \frac{u'}{r} - \frac{2 u (f^2 + g^2)}{1+q(f^2+g^2)} + \nn \\ && \:\:\:+\frac{2 f^2}{1+q(f^2+g^2)}  =  0 \label{EmDilNOu} \\
&& g'' + \frac{g'}{r} - \frac{u^2 g}{r^2} - \frac{q}{2}  \left(\frac{u'}{r} \right)^2 g  -\nn\\
  &&\:\:\:-\beta(f^2 + g^2 -1)g  =  0 \label{EmDilNOf} \\
&& f'' + \frac{f'}{r} - \frac{f}{r^2}(1-u)^2 - \frac{q}{2} \left(\frac{u'}{r} \right)^2 f - \nn \\
&& \:\:\: -  \beta(f^2 + g^2 -1)f  =  0 \label{EmDilNOf}
\eear
Clearly the NO ansatz with $g=0$ is an ``embedded" solution to these equations.\\
We assume a  dilatonic coupling of the form
\be
B(\Phi^\dag \Phi)=1+\frac{q\; \Phi^\dag \Phi}{\eta^2}
\ee
\\
and we can write the energy of the vortex as
\bear
E & = & \int^{\infty}_0 d r~r \: (g'^2 + f'^2 + \frac{f^2}{r^2}(1-u)^2 + \frac{u^2 g^2}{r^2}  \nn \\
&& \: + \frac{1}{2}(1+q(f^2 +g^2))\left(\frac{u'}{r} \right)^2 \nn\\
&& \:+ \frac{\beta}{2}(f^2 + g^2 -1)^2) \label{embdilNOen} \\
& = & E_0 + \delta E_g
\eear
where
\bear
E_0 & = & \int^{\infty}_0 d r \: (f'^2 + \frac{f^2}{r^2}(1-u)^2 + \frac{1}{2}(1+ q f^2)\left(\frac{u'}{r} \right)^2 \nn \\
&& \: +\frac{\beta}{2}(f^2  -1)^2)
\eear
is the unperturbed energy and the energy perturbation due to $g$ can be written as
\be
\delta E_g  =  \int^{\infty}_0 d r \: r (g \hat{O} g)
\ee
where $\hat{O}$ is a Schrodinger-like Hermitian operator of the form:
\bear
\hat{O} & = & -\frac{1}{r} \frac{d}{dr}\left(r\frac{d}{dr} \right) + \frac{u^2}{r^2} + \frac{q}{2}\left(\frac{u'}{r} \right)^2 \nn \\ && \:+  \beta (f^2 -1)
\eear
The Schrodinger potential of $\hat{O}$ is
\bear
V_{\it Schrodinger} & = & \frac{u^2}{r^2} + \frac{q}{2}\left(\frac{u'}{r} \right)^2 +  \beta (f^2 -1)
\eear
For values of the parameters $q$ and $\beta$ for which $\hat{O}$ has no negative eigenvalues we have $\delta E_g \geq 0$ and therefore no instability develops. The plot in Fig. \ref{fig:stab_power_law.eps} shows the increase of the stability region as we increase the value of the parameter $q$. Note that the plot is not identical to that shown in Ref. \cite{Perivolaropoulos:2013uea} (Fig. \ref{fig:stab_exp_law.eps}) since the functional form of the dilatonic coupling is different in our case (we assume a power law while an exponential $B(\Phidag \Phi)=e^{\frac{q \Phidag \Phi}{\eta^2}}$ was assumed in \cite{Perivolaropoulos:2013uea}).
\begin{figure}[!ht]
\centering
\includegraphics[scale=0.6]{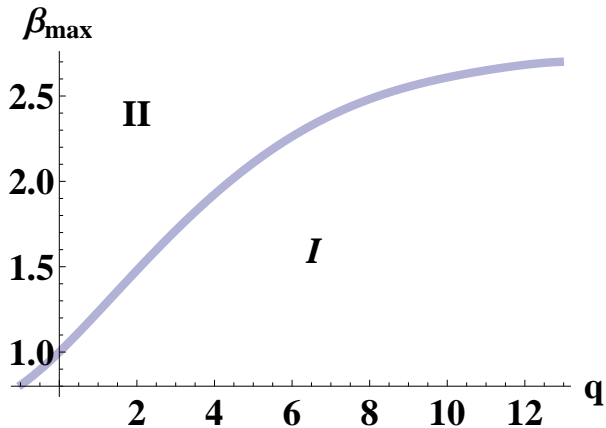}
\caption{The stability region $\beta(q)$ for the embedded dilatonic NO vortex. In this case we have assumed a power law dilatonic function. Sector $I$ is the stability region while sector $II$ is the instability region. Notice that for $\beta<1$ a negative $q$ can destabilize the vortex. In that region the instability leads to a stable scalar field condensate confined in the core.}
\label{fig:stab_power_law.eps}
\end{figure}

\begin{figure}[!ht]
\centering
\includegraphics[scale=0.5]{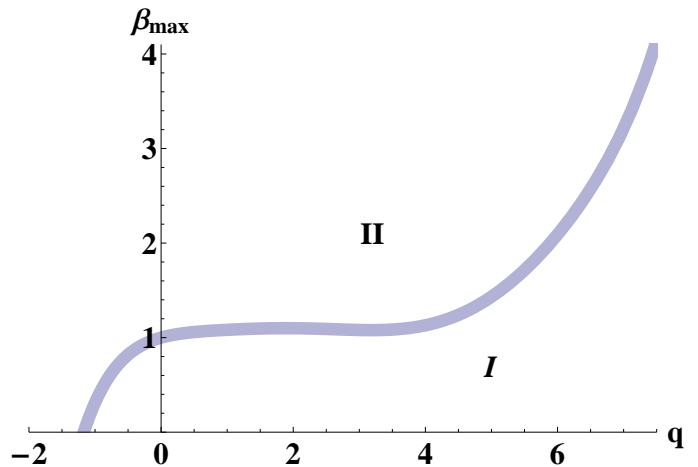}
\caption{The stability region $\beta(q)$ for the embedded dilatonic NO vortex. An exponential law ($B(\Phidag \Phi)= e^{\frac{q \Phidag \Phi}{\eta^2}}$) is implemented for the dilatonic function (see also \cite{Perivolaropoulos:2013uea}) but this time negative values of q are included. Sector $I$ is the stability region while sector $II$ is the instability region. For $\beta<1$ a negative $q$ can destabilize the vortex and lead to the formation of a scalar field condensate that does not propagate outside the core.}
\label{fig:stab_exp_law.eps}
\end{figure}
Alternatively, the stability analysis may be performed at the nonlinear level by performing a full-minimization of the energy (\ref{embdilNOen}) with respect to $f,u,g$. In order to achieve this we used a simple mathematica code where we performed the minimization with fixed NO boundary conditions. In Fig. \ref{fig:Fields(b=1.5, q=0)} the $f,u,g$ fields are shown for $\beta=1.5$ and $q=0$. As expected, instability clearly develops in this case since the perturbation grows in the core while the NO vortex core size increases out to the boundary where it is artificially confined by the boundary conditions.
\begin{figure}[!ht]
\centering
\includegraphics[scale=0.4]{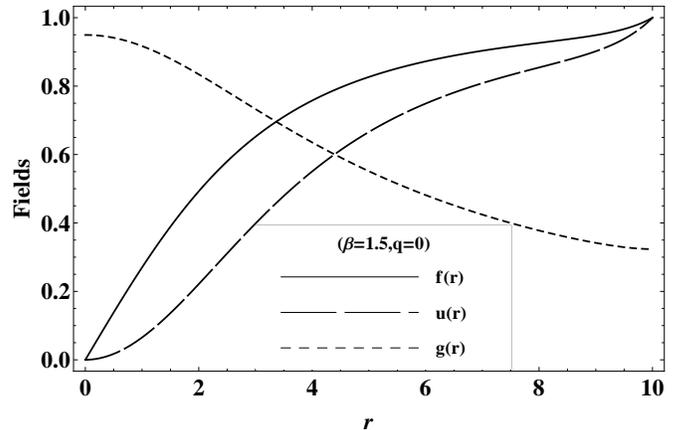}
\caption{Solutions for $f(r)$, $u(r)$ and $g(r)$ for the embedded dilatonic NO vortex for $\beta=1.5$ and $q=0$. We notice that, as expected, for $\beta>1$ the NO vortex is unstable to perturbations orthogonal to $\Phi_2$. This means that above a critical value for $\beta$, the scalar field $\Phi$ chooses to develop a component g(r)(dashed line) towards the z-direction in order to reduce the increased potential energy.}
\label{fig:Fields(b=1.5, q=0)}
\end{figure}

However, as the value of $q$ increases, the value of $\beta_{eff}$ decreases and we have a stability improvement. This is demonstrated in Fig. \ref{fig:Fields(b=1.5, q=5.5)} where it is shown that after energy minimization, no instability develops for $q=5.5$ and the same value of $\beta=1.5$. Indeed, as we increase the value of $q$ it becomes more costly energetically for the scalar field to develop a non-zero value at the defect core (due to the term $\frac{1}{2}(1+q(f^2 +g^2))\left(\frac{u'}{r} \right)^2$) where the gauge field kinetic term is non-zero and positive definite.

\begin{figure}[!ht]
\centering
\includegraphics[scale=0.5]{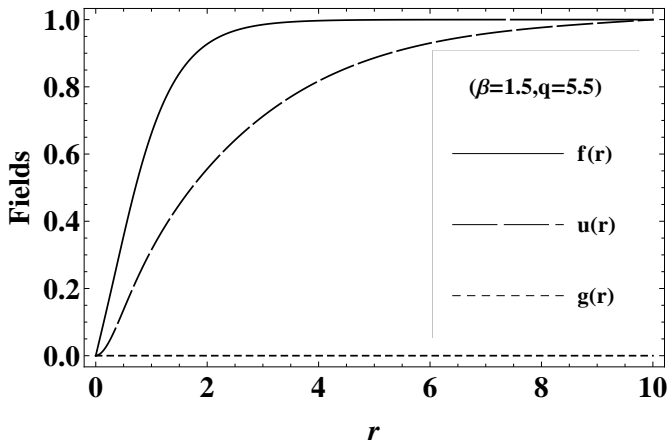}
\caption{Solutions for $f(r)$, $u(r)$ and $g(r)$ for the embedded dilatonic NO vortex for $\beta=1.5$ and $q=5.5$. We notice that for a non-zero value of q we can have vortex solutions for $\beta>1$.}
\label{fig:Fields(b=1.5, q=5.5)}
\end{figure}

Finally we point out an interesting effect that occurs for negative values of $q$ and $\beta<1$. In this case $\beta_{eff}$ may become larger than 1 in the core region while away from the core $\beta_{eff}<1$. Thus the instability develops inside the core but does not propagate outside where the stability condition holds. We have verified numerically by energy minimization, the existence of such a localized scalar field condensate. The corresponding field configuration is shown in Fig. \ref{fig:condensate}

\begin{figure}[!ht]
\centering
\includegraphics[scale=0.41]{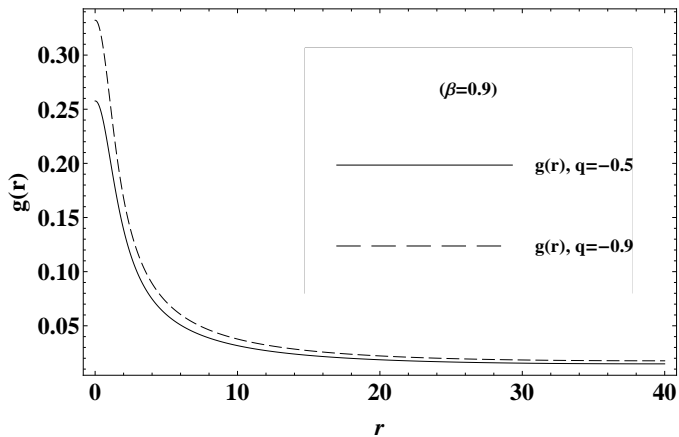}
\caption{Solutions for $g(r)$ for the embedded dilatonic NO vortex for $\beta=0.9$ and $q=-0.9, q=-0.5$. Notice that $g$ develops a non-zero value in the core of the defect which remains localized there as the energy is minimized. A localized scalar field condensate forms. }
\label{fig:condensate}
\end{figure}

\subsection{Global Vortex in an External Gauge Field}

Consider now the global field Lagrangian
\be
\mathscr{L}= (\D_{\mu} \Phi)^{\dag} \D^{\mu} \Phi -\frac{1}{4} B(\Phi^\dag \Phi) F_{\mu \nu} F^{\mu \nu} - V(\Phi^\dag \Phi) \label{EmbDilGl}
\ee
where the gauge $U(1)$ symmetry has been replaced by a global $SU(2)$, while keeping the gauge field kinetic term and
\be
V(\Phi^\dag \Phi)=\frac{\lambda}{4}(\Phi^\dag \Phi -\eta^2)^2
\ee
Using now the embedded vortex ansatz
\be
\Phi=
\left(
\begin{array}{c}
\delta\Phi_1 \equiv g \\
f e^{i \theta}\\
\end{array} \right) \label{doublet}
\ee
we obtain the field equations
\bear
&& g'' + \frac{g'}{r} - \frac{q}{2}B_z^{~2} g - (f^2 + g^2 -1)\frac{g}{2} = 0 \label{DilGlVoreq1} \\
&& f'' + \frac{f'}{r} - \frac{f}{r^2} - \frac{q}{2}B_z^{~2} f- (f^2 + g^2 -1)\frac{f}{2} = 0 \label{DilGlVoreq2}
\eear
The corresponding energy density is
\bear
\rho & = & g'^2 + f'^2 + \frac{f^2}{r^2} + \frac{1}{2} (1+q(f^2+g^2)) B_z^{~2} \nn \\
&& \:+ \frac{1}{4}(f^2 + g^2 - 1)^2 \label{DilGlVorEn}
\eear
where the following rescaling has been used
\bear
&& f \to \bar{f}  =  \eta f \\
&& g \to \bar{g}  =  \eta g \\
&& r \to \bar{r} = \frac{r}{\eta \sqrt{\lambda}} \\
&& B_z \to \bar{B_{z}} = B_z \eta^2 \sqrt{\lambda}
\eear
We assume a {\it{Gaussian}} for the external magnetic field of the form
\be
B_z= B_{z0}~ e^{-\frac{r^2}{r_0^2}}
\ee
In order to investigate the stability of the embedded global vortex we first fully minimize the energy with respect to the $f,g$ fields with the proper boundary conditions.
\\
Using the NO boundary conditions it is trivial to obtain the solution $g(r)$ for various values of the parameters $B_{z0},r_0$. The energy to be minimized is of the form
\bear
E & = & \int dr~r (f'^2 + g'^2 +\frac{f^2}{r^2} \nn \\
 &&\:+ \frac{1}{2}(1+q (f^2 + g^2)) (B_{z0}^{~2}\: e^{\frac{-r^2}{r_{0}^{2}}})^2 \nn \\
 && \:+ \frac{1}{4}(f^2+g^2 -1)^2 ) \label{DilGlVoPerEn}
\eear
In Fig. \ref{fig:g(r)stable} we show the form of $g(r)$ after energy minimization with fixed boundary for various values of $B_{z0}$ and $r_0$. For large magnetic field magnitude, the development of non zero field $g(r)$ in the core where the gauge field is excited, becomes energetically costly and therefore no instability develops there. Away from the core however the instability remains due to the global nature of the symmetry and the limited range ($r_0$) of the magnetic field.

\begin{figure}[!ht]
\centering
\includegraphics[scale=0.37]{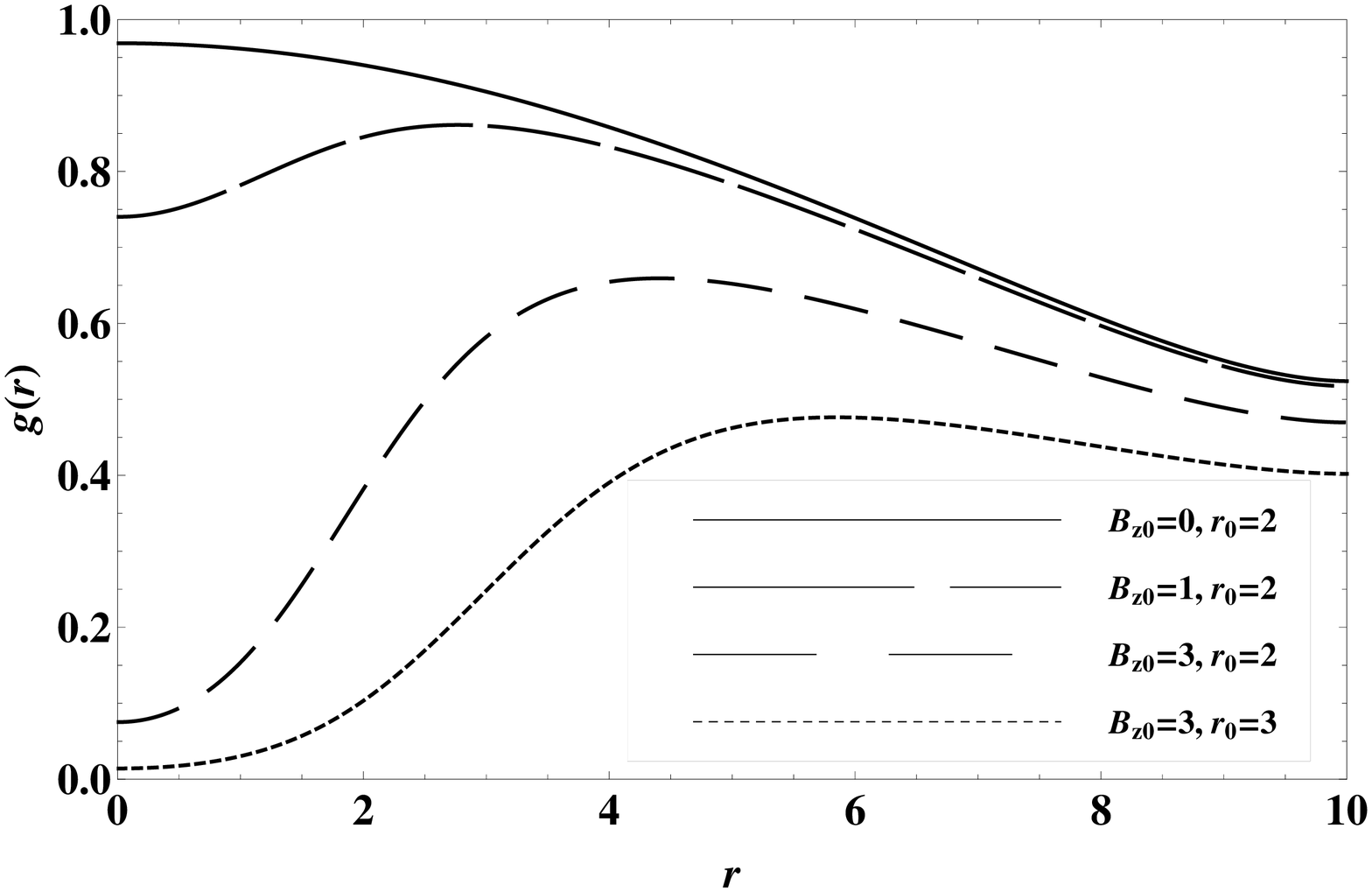}
\caption{The form of $g(r)$ after energy minimization with fixed boundary for various values of $B_{z0}$ and $r_0$.}
\label{fig:g(r)stable}
\end{figure}
Another interesting point would be to examine the changes, the insertion of a dilatonic coupling, induces to the behavior of the unperturbed field $f(r)$. The term $\frac{1}{2}(1+q
f^2) (B_{z0}^2\: e^{\frac{-r^2}{r_{0}^{2}}})^2$ of the unperturbed energy changes the effective potential from $\frac{1}{4}(f^2-1)$ (by regrouping the terms of the unperturbed energy to complete squares) to
\be
V_{eff}=\frac{1}{4}(f^2-(1-q B_z^{~2}))^2 \label{effectivepot}
\ee
where we observe that a negative $q$ forces $f$ to increase in the region where $B_z$ is non-zero (near the core). This behavior of $f$ is depicted on Fig (\ref{fig:f(r)local}). As expected for $r\to \infty$, $B_z \to 0$ and we end up to the initial potential.
\begin{figure}[!ht]
\centering
\includegraphics[scale=0.34]{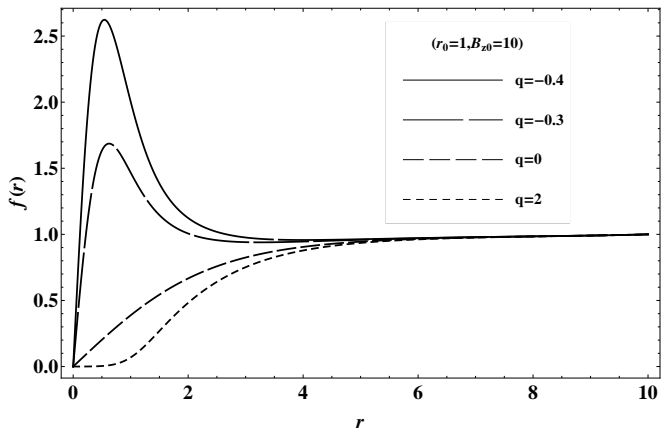}
\caption{Solution of the unperturbed field $f(r)$ for the embedded dilatonic NO Vortex for various values of $q$. We observe that $f(r)$ rapidly increases at the region where the magnetic field is significant. This suggests a change of the effective potential. It also suggests that the scalar field $f(r)$ prefers energetically to adopt a high kinetic energy in order to reduce $V_{eff}$ at that area.}
\label{fig:f(r)local}
\end{figure}

It is also straightforward to verify the above stability results considering small perturbations on the embedded global vortex.  The energy (\ref{DilGlVoPerEn}) may be expressed as the sum of the unperturbed energy and the energy perturbation due to $g$ as
\bear
E & = & \int dr~r(f'^2 + \frac{f^2}{r^2}+ \nn\\
 &&\:+ \frac{1}{2}(1+q f^2)(B_{z0}\: e^{\frac{-r^2}{r_{0}^{2}}})^2 \nn \\
  &&\:+ \frac{1}{4}(f^2 -1)^2)   +\:\delta E_g
\eear
where the energy perturbation due to $g$ can be written as
\be
\delta E_g = \int^{\infty}_0 dr~ r (g \hat{O} g)
\ee
where $\hat{O}$ is a Schroedinger-like Hermitian operator of the form
\be
\hat{O} = - \frac{d^2}{d r^2} - \frac{1}{r} \frac{d}{dr} + \frac{q}{2} (B_{z0} \: e^{\frac{-r^2}{r_{0}^{2}}})^2 + \frac{1}{2} (f^2 -1) \label{Ohat}
\ee
and we have kept only terms up to second order in $g$.\\
Also the Schrodinger potential corresponding to $\hat{O}$ is
\bear
V_{\it{Schrodinger}} & = & \frac{q}{2} (B_{z0}\: e^{\frac{-r^2}{r_{0}^{2}}})^2 + \frac{1}{2} (f^2 -1) \label{DilGlVorVS}
\eear
The embedded dilatonic global vortex is stable for the parameter range for which the operator $\hat O$ has no negative eigenvalues. The existence of negative eigenvalues depends on the depth of the Schrodinger potential (\ref{DilGlVorVS}) shown in Fig. \ref{fig:v(q)}, for three values of the parameter $q$ and fixed $B_{z0}$ and $r_0$. Clearly for positive values of $q$ the potential becomes more repulsive for $r<r_0$. However, for $r>>r_0$ the potential remains unaffected. Due to the assumed limited range of the external magnetic field the operator $\hat O$ is found to have negative eigenvalues for any finite value of the parameters $q$,$B_{z0}$ and $r_0$. This is demonstrated in Fig. \ref{fig:g(q)} which shows the solution of the equation
\be
{\hat O} g(r) = 0
\label{hato0}
\ee
with boundary conditions $g(0)=1$, $g'(0)=1$. For $r<r_0$ where the potential is repulsive, the ``candidate'' zero mode increases with $r$ but eventually, for large $r$ the attractive nature of the potential dominates and we get $g(r)<0$ indicating the existence of negative eigenvalues. This behavior of $g(r)$ may be interpreted in accordance with the above energy minimization result, as localized stability of the dilatonic embedded vortex in the region of the external magnetic field.

\begin{figure}[!ht]
\centering
\includegraphics[scale=0.33]{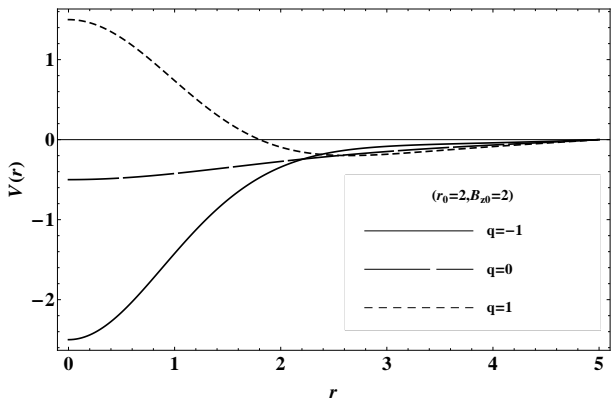}
\caption{The Schroedinger potential of the operator $\hat O$ (\ref{Ohat}) becomes repulsive for $q>0$ and $r<r_0$.}
\label{fig:v(q)}
\end{figure}

\begin{figure}[!ht]
\centering
\includegraphics[scale=0.33]{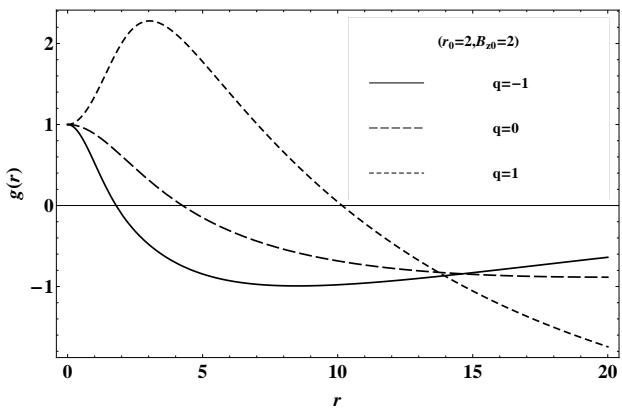}
\caption{The solution of the equation ${\hat O} g(r) = 0$ (\ref{hato0}) for different values of the parameter $q$. The solution increases with $r$ for $r\approx r_0$ when the potential is repulsive ($q=1$). However, eventually the solution, ends up negative for large $r$ leading to instability. This behavior of $g(r)$ confirms the results obtained from the energy minimization method.}
\label{fig:g(q)}
\end{figure}

\section{Monopoles}

\subsection{Dilatonic 't Hooft-Polyakov Monopole}

Monopoles form in field theories involving symmetry breaking phase
transitions where the vacuum manifold is $M \cong{S^2}$
\cite{Preskill:1992bf}. This is the case for example when an
$SO(3)$ symmetry gets spontaneously broken to $U(1)$.

Consider for example the Lagrangian density describing an $O(3)
\rightarrow O(2)$ symmetry breaking which accepts magnetic 't
Hooft-Polyakov monopole solutions
\cite{'tHooft:1974qc} with a
dilatonic coupling, \be \mathcal{L}= \half\,\left(D_\mu\,\Phi^a
\right)\left(D^\mu\,\Phi^a\right) - V(\Phi^a) -\frac{B(\Phi^a)}{4
e_0^2}\,F_{\mu\nu}^{a}\,F^{a\mu\nu}, \label{laga} \ee where $
B(\Phi^a)$ describes a possible variation of the gauge charge (and
thus for the effective charge we have $e^2=e_0^2/B(\Phi^a)$)
where $a=1,2,3$ are internal indices. As usual we
define the non-Abelian gauge field strength by \be
F_{\mu\nu}^a=\partial_\mu A_\nu^a -
\partial_\nu A_\mu^a + e_0 \epsilon^{a\,b\,c}\,A_{\mu}^b
A_{\nu}^c, \label{gfs} \ee

The covariant derivatives are written
in the usual form \be D_\mu \Phi^a = \partial_\mu \Phi^a +
e_0\,\epsilon^{a\,b\,c} a_\mu^b\,\Phi^c, \label{cd} \ee
 where $\epsilon^{a\,b\,c}$ is the Levi-Civita tensor.

The symmetry breaking potential $V(\Phi^a)$ is of the usual form \be
V(\Phi^a) =
\frac{\lambda}{4}\,\left(\Phi^a\,\Phi^a-\eta^2\right)^2 \ee

The 't Hooft-Polyakov monopole ansatz \cite{'tHooft:1974qc} is of the form \bear
\Phi^a(r)&=&X(r)\,\frac{x^a}{r},\label{Xig}\\
a_0^a(r)&=&0,\label{a0}\\
a_i^a(r)&=&\epsilon_{iak}\,\frac{x_k}{e_0\,r^2}\,[W(r)-1],\label{ai}
\eear where $x^a$ are the Cartesian coordinates and
$r^2=x^k\,x_k$. $X(r)$ and $W(r)$ are radial functions obtained by
minimization of the self-energy, i.e., the mass of the monopole
\be\ E = 4\,\pi\int_0^\infty dr \; r^2 \; \rho\label{degvs}\ee or
by solving the field equations. The energy
density is obtained from the Lagrangian (\ref{laga}) as
\be\rho=T_{00}=-g_{00}\mathcal{L}\ee

After a rescaling of the form
\bear
X \to {\bar X} & = & \eta X \label{rescalef} \\
r \to \bar{r} & = & \frac{r}{\eta e_0}\label{rescaler} \eear
the energy density becomes
\bear \rho &=& \frac{\eta}{e_0} \bigg[ B(X)\left[\left(\frac{W'}{r}
\right)^2 + \frac{1}{2} \left(\frac{1-W^2}{r^2}\right)^2 \right]+
\frac{(X')^{2}}{2}\nn\\ &+& \left(\frac{W X}{r} \right)^2+
\frac{\beta}{2}(1-X^2)^2 \bigg] \label{ro}\eear
 with a prime meaning a derivative with respect to the dimensionless coordinate $r$.
 Using (\ref{degvs}) we obtain:

 \bear \ E &=&\frac{4\pi\eta}{e_0}\int_{0}^{\infty}dr\{\frac{r^2}{2}\left(\frac{dX}{dr}\right)^2+X^2 W^2+
\frac{\beta
r^2}{2}(1-X^2)^2\nn\\&+&B(X)\left[\left(\frac{dW}{dr}\right)^2+\frac{(1-W^2)^2}{2r^2}\right]\}\label{act}\eear
The dimensionless parameter $\beta$ is defined as in the case of vortices as
$\beta \equiv (\frac{m_\Phi}{m_A})^2$  where
$m_\Phi=\frac{\sqrt{\lambda} \eta}{\sqrt{2}}$ and $m_A= e_0\,
\eta$ are the masses of the scalar and gauge fields respectively.
Thus $\beta = \frac{\lambda}{2 e_0^2}$.

In order to find the solution to the field equations, we minimize the energy (\ref{act}) using the boundary
conditions: $X(r\rightarrow\infty)=1$,
$X(r\rightarrow 0)=0$,
$W(r\rightarrow\infty)=0$ and $W(r\rightarrow
0)=1$.

\begin{figure}[!ht]
\centering
\includegraphics[scale=0.8]{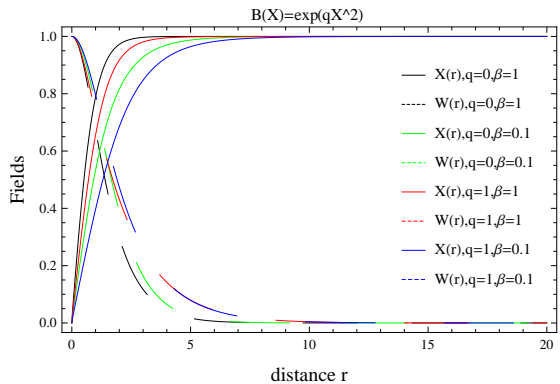}
\caption{Solutions for $X(r)$, $W(r)$ for the dilatonic magnetic
monopole when $ B(X)= e^{q X^2}$ for several values of the
parameters $\beta$ and $q$. The same color in the fields
corresponds to the same values of $\beta$ and $q$. Notice that as
$\beta$ increases, the slope of the curves increases and the
fields acquire their vacuum expectation values for smaller $r$.}
\label{fig:dilexp}
\end{figure}

 Without loss of generality we normalize $e$ so that $B(r\rightarrow 0)=1$.
In this section we parametrize the dilatonic coupling as
 \be\ B(X)= e^{q X^2}\label{ansatz1}\ee

In order to obtain the dilatonic 't Hooft-Polyakov monopole solution, we minimize the energy (\ref{act}) using the above boundary
conditions, for several values for the parameters $\beta$ and $q$. In Fig. \ref{fig:dilexp} we show the resulting fields $X(r)$ and
 $W(r)$ when $(q=0,\beta=0.1)$,$(q=0,\beta=1)$,$(q=1,\beta=0.1)$,
 $(q=1,\beta=1)$. For each pair of the fields $X(r)$ and $W(r)$ we
 use the same color for the plot in order to be easily
 visible. As in the case of dilatonic vortices, decreased value of $\beta$ and increased value of $q$ leads to a dilatonic monopole with larger core scale.

We have verified that a polynomial form of $B(X)$ (i.e. $B(X)= 1+q
 X^2$ leads to similar results (see Fig. \ref{fig:dilqua}).

\begin{figure}[!ht]
\centering
\includegraphics[scale=0.8]{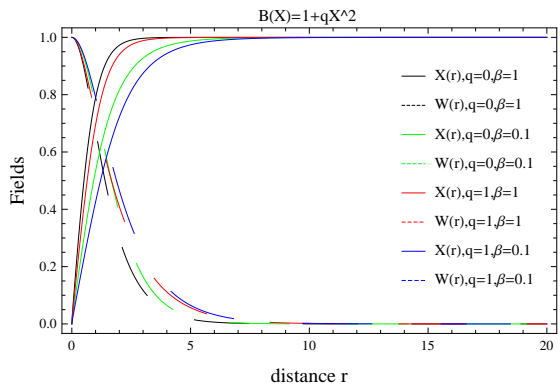}
\caption{Solutions for $X(r)$ and $W(r)$ for the dilatonic
magnetic monopole when $B(X)=1+qX^2$ for the same values of the
parameters $\beta$ and $q$ as in Fig. \ref{fig:dilexp}. The same color in the
fields corresponds to the same values of $\beta$ and $q$.}
\label{fig:dilqua}
\end{figure}

\subsection{Embedded Dilatonic Monopole}

We now consider the
embedding of the gauge monopole \cite{'tHooft:1974qc} in a model with $O(4)$ symmetry\cite{Barriola:1993fy,Gibbons:1992gt}. This is
achieved by adding in the  scalar $\Phi$ one more component as \be
\Phi^4(r)=g(r) \label{ggen} \ee  The embedded monopole potential
(semilocal monopole) takes the form \be V(\Phi^a) = \frac{\lambda}{4}\,\left(X(r)^2
+g(r)^2-\eta^2\right)^2 \ee Using the methods and arguments of
Ref. \cite{Barriola:1993fy} it is straightforward to show that the embedded
dilatonic monopole solution in this model is unstable for all
values of parameters. The instability persists because the embedded gauge group $O(3)$
acts trivially on the additional field component $\Phi^4(r)=g(r)$.
In this case it may be shown that there is a smooth sequence
of field configurations parametrized by a parameter $\xi$ with
energy monotonically decreasing with $\xi$ that starts from the
embedded monopole configuration for $\xi=0$ and ends at the
vacuum for $\xi=\pi/2$. In view of this
simple and powerful result we omit presenting the perturbative
energy mininization analysis of the embedded dilatonic monopole
which involves minimization of the embedded gauged monopole energy
corresponding to the energy density \bear \rho &=&
\frac{\eta}{e_0}\bigg[B(X)\left[\left(\frac{W'}{r} \right)^2 +
\frac12 \left(\frac{1-W^2}{r^2}\right)^2 \right]+
\frac{(X')^{2}}{2}+ \nn\\ &+&\left(\frac{WX}{r} \right)^2+
\frac{(g')^{2}}{2}+\frac{\beta}{2}(1-X^2-g^2)^2 \bigg]
\label{ro1}\eear Such an analysis simply verifies the anticipated
instability for all values of the parameters $\beta$ and
$q$. The corresponding
analysis for the embedded dilatonic global monopole also leads to
instability either using the approach of Ref. \cite{Barriola:1993fy} or through
direct energy minimization of the density \bear \rho &=&
\frac{\eta}{e_0}\Big[
B(X)(e^{-\frac{r^2}{r_{0}^2}})^2+ \frac{(X')^{2}}{2}+ \nn\\
&+&\left(\frac{X}{r} \right)^2+
\frac{(g')^{2}}{2}+\frac{\beta}{2}(1-X^2-g^2)^2 \Big]
\label{globalro1}\eear where we have assumed a similar external
gauge field as in the previous
section. The field
configurations that minimize the above energy using boundary
conditions at $r=10$ ($X(r=10)=1$) are shown in Fig.
\ref{fig:Xqqglobal}

\begin{figure}[!ht]
\centering
\includegraphics[scale=0.8]{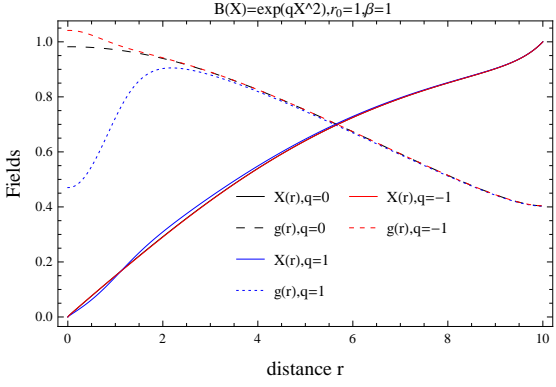}
\caption{The energy minimizing fields $g(r)$ and $X(r)$ as a function of distance
$r$ for embedded global dilatonic monopole, in the presence of
a gaussian external ``magnetic'' field $e^{-\frac{r^2}{r_{0}^2}}$
for several values of the parameter $q$, when $\beta=1$ and $r_0=2$. Here we assumed an exponential dilatonic coupling $B(X)= e^{q X^2}$.}
\label{fig:Xqqglobal}
\end{figure}

Notice that as expected the instability tends to expand outwards
leading to the vacuum in regions away from the external field
region. However, for $r<r_0$ where the external field is significant,
the field remains out of the vacuum due to the effects of the
external field which stabilizes locally the embedded global
monopole as in the case of the embedded global vortex discussed in
the previous section.

\section{Physical Effects}

The NO vortex can be embedded in various other generalizations of the Abelian-Higgs model. One of the most interesting cases, is the bosonic sector of the standard Glashow-Salam-Weinberg (GSW) electroweak model.
This bosonic sector
corresponds to a symmetry breaking $SU(2)_L\times U(1)_Y$ model with a scalar field
$\Phi$ in the fundamental representation of $SU(2)_L$.  Assuming in addition a dilatonic coupling, the Lagrangian takes the form:
\begin{eqnarray}
{\cal L} &=&  -\frac{1}{4} B(\Phidag \Phi) W^{a}_{\mu
\nu}W^{a\mu \nu} -
\frac{ 1}{4 } B(\Phidag \Phi) Y_{\mu \nu}Y^{\mu \nu}
\nonumber\\
& + & (D_{\mu}\Phi)^\dag D^{\mu}\Phi - \frac{\lambda}{4} (\Phidag \Phi - \eta^2)^2
\end{eqnarray}
where $B(\Phidag \Phi)$ represents the dilatonic coupling while $\ W_{\mu \nu}^{\ \ a} = \partial_\mu W_\nu^a - \partial_\nu W_\nu^a + g
\epsilon^{abc} W_\mu^b W_\nu^c \
$
and $\ Y_{\mu \nu}  = \partial_\mu Y_\nu - \partial_\nu Y_\mu \ $ are the field
strengths for the $SU(2)_L$ and $U(1)_Y$ gauge fields  respectively. Also $D_{\mu}=\partial _\mu -{ig}  \tau ^a W_\mu ^a -
{ig'}  Y_\mu$ is the covariant derivative and $g$, $g'$ are the two gauge couplings ($\tau^a$ are the Pauli matrices).

In the unitary gauge, the $Z$ and $A$ fields
are defined as
\begin{equation}
Z_\mu \equiv \cos\theta_wW_\mu^3 - \sin\theta_w Y_\mu \ ,
\ \ \ \
A_\mu \equiv \sin\theta_wW_\mu^3+ \cos\theta_w Y_\mu \ ,
\label{2.9}
\end{equation}
and $W_\mu^\pm \equiv (W_\mu^1 \mp i W_\mu^2) / \sqrt 2 $ are the W bosons.
The weak mixing angle $\theta_w$ is given by $\tan\theta_w \equiv {{g'} / g} $;
electric charge is
$e = g_z \sin\theta_w \cos\theta_w$ with $g_z \equiv (g^2 +
{g'} ^2 )^{1/2} $. After proper rescaling the only two free parameters of the bosonic sector become the weak mixing angle and the parameter $\beta\equiv m_H^2 / m_z^2$ defined as the ratio between the Higgs mass $m_H=2\lambda \eta$ over the Z-boson mass $m_Z=g_z \eta /2$.

The simplest embedding of the NO string in the dilatonic electroweak model  corresponds to a Z-string along the $z$-axis described as
\cite{Vachaspati:1992jk}:
\be
\Phi= \left(
\begin{array}{c}
\Phi_1 \\
\Phi_2\\
\end{array} \right) =
\left(
\begin{array}{c}
0 \\
f e^{i \theta}\\
\end{array} \right) \label{zstring1}
\ee
\bear
Z_\mu &=& \frac{u(r)}{r}{\hat \theta} \label{zstring2} \\
A_\mu &=& W_\mu^\pm = 0
\label{zstring3}
\eear
where $f$ and $u$ are the dilatonic NO solutions.  It is straightforward to show that this is a
solution of the dilatonic GSW bosonic sector equations of motion (or equivalently an extremum of the GSW dilatonic bosonic sector energy).
However, like the dilatonic semilocal string, the dilatonic electroweak Z string is unstable and can decay
by unwinding to the vacuum.

The solution (\ref{zstring1})-(\ref{zstring3}) and the electroweak bosonic sector reduce to the semilocal string model in the limit
$\sin^2\theta_w = 1$ and therefore in this limit the electroweak string is classically stable for
$\beta < 1$ and unstable for $\beta > 1$. For $\sin^2\theta_w < 1$ the stability of the electroweak string persists but for a smaller range of the parameter $\beta$ \cite{James:1992wb}. When there is no dilatonic coupling, the physical values of the parameters $\beta$ and $\theta_w$ correspond to unstable electroweak string.

In the presence of a large enough dilatonic coupling, the corresponding stability region increases arbitrarily as discussed in section II. We anticipate that this stability improvement will persist even for the experimentally measured parameter values $\theta_w$ and $\beta$. It is therefore important to identify the required value of the dilatonic coupling $q$ for stability of the dilatonic electroweak string for the measured values of $\theta_w$ and $\beta$. We postpone this analysis for a later publication but we point out that if the required value of $q$ for stability is consistent with current experiments, then the possibility of formation of metastable dilatonic electroweak strings in accelerators and/or in the early universe arises. In this case we anticipate the existence of interesting signatures and effects in both accelerator and cosmological setups. In particular such effects include:
\begin{itemize}
\item
{\bf Primordial Magnetic Fields:}  A gas of metastable electroweak segments formed during the electroweak phase transition is necessarily accompanied
by a gas of electroweak monopoles. The eventual collapse and disappearance of electroweak
strings removes all the electroweak monopoles but the long range magnetic field emanating
from the monopoles is expected to remain trapped in the cosmological plasma. This will then lead to a residual primordial magnetic field
in the present universe. An estimate of the average flux of this primordial magnetic field was obtained in \cite{Vac91,Vac94}.
\item
{\bf Generation of Baryon Number-Cosmic Rays:}
A gas of metastable electroweak string segments and loops would, in general, contain
some helicity density of the Z-field. So when the electroweak strings eventually annihilate,
it is possible that the helicity gets converted into baryon number \cite{Vac94,VacFie94}. In more exotic models (such as this), strings at the electroweak
scale that were stable and had superconducting properties, could also be
responsible for baryogenesis \cite{Bar95} and the presence of primary antiprotons in cosmic rays \cite{StaVac96}.
\item
{\bf Variation of Fine Structure Constant:} A dilatonic coupling in models involing electromagnetism like the electroweak model, leads naturally to the possibility of variation of the fine structure constant $\alpha$. In the presence of a metastable dilatonic electroweak string this variation is anticipated to be spatial on the scale of the core of the dilatonic defect. Such microscopic localized variation of $alpha$ could be detectable in accelerators where either metastable dilatonic electroweak strings or dilaton-Higgs particles are produced and decay. This effect becomes more interesting in view of the recent claim for a $4\sigma$ detection of spatial variation of $\alpha$ on cosmological scales obtained from careful analysis of quasar absorption spectra \cite{King:2012id}.
\item
{\bf Signatures in accelerators: Dilatonic Dumbells} The production of solitonic states in particle accelerators as well as their experimental signatures constitute open issues that become particularly important in the context of the existence of metastable electroweak strings. A rotating electroweak monopole-antimonopole pair connected by a Z-string and stabilized by a centrifugal barrier is known as a dumbell. The decay signature of such a metastable system was first studied by Nambu \cite{Nambu:1977ag} who estimated the energy and angular momentum of such a system as well as its lifetime and decay products. In the context of a dilatonic coupling such a system may get stabilized not only due to its angular momentum but also at the field theoretic level. Thus we anticipate an increased lifetime and a cleaner signature in accelerators.

\end{itemize}

Thus, the role of electroweak strings in cosmology depends on their abundance during and after
the electroweak phase transition. If this abundance is negligible, electroweak strings may at
best only be relevant in future accelerator experiments. Such relevance is expected to increase significantly in the presence of a dilatonic coupling which is anticipated to improve their stability.

\section{Conclusions-Discussion}
Topological defects formed in theories where the scalar field
couples to the gauge field strength tensor (dilatonic defects)
have significant novel properties. In
particular
\begin{itemize}
\item Their core scales can be significantly larger than the
corresponding ordinary defects with minimal
coupling. \item The
corresponding embedded defects have modified stability properties.
\item The instability of global dilatonic defects in the presence
of an external gauge field does not proceed towards the vacuum.
Instead it proceeds towards a field configuration which deviates
from the vacuum in the region where the external gauge field is
excited. This
configuration may be interpreted as a local stabilization of the
global embedded defects \item The instability of the gauged
embedded vortex may proceed (for certain parameter values) towards
a scalar field condensate where the instability is excited but is
confined to the region of the embedded defect core.
\end{itemize}
Interesting extensions of this analysis are the following:
\begin{itemize}
\item Investigation of the stability properties of more realistic
embeddings like the electroweak vortex\cite{Achucarro:1999it,James:1992zp,Vachaspati:1992jk,Nambu:1977ag}. The
possible formation of new electroweak vortices with core
condensates representing confinement of the instability is a
particularly interesting prospect. In
addition our analysis hints towards the possible stabilization of
the electroweak vortex for realistic parameter values in the
presence of a dilatonic coupling. \item
Investigation of the core properties of other dilatonic defects
like textures, skyrmions and domain walls in the presence of
external gauge fields. In this case, there is the possibility of formation of condensates
similar to the one found for the dilatonic gauged vortex in the
parameter region where the embedded defect is stable but gets
destabilized due to the dilatonic coupling. This mechanism for the
formation of condensates from embedded defects appears to be
generic in the context of dilatonic embedded defects. \item The
presence of dilatonic defects with Hubble scale cores could
naturally induce spatial variation of the corresponding gauge
charges and in particular of the fine structure constant. This
prospect is interesting in view of the recent claims of the
existence of a fine structure constant dipole on cosmological
scales obtained from the absorption spectra of quasars on
cosmological scales \cite{King:2012id,Mariano:2012wx}. This class of models
naturally predicts an alignment of the fine structure constant and
dark energy dipoles. Indications for such an alignment have been
observed recently in a combination of type Ia data and quasar
absorption spectra data \cite{Mariano:2012wx}. Thus, the detailed study of the
cosmological properties of this class of models (extended
topological quintessence) constitutes an exciting extension of the
present analysis. \item A systematic review of
experimental/observational constraints on the parameter $q$ and
the dilatonic coupling in realistic theories (eg in extensions of
the standard electroweak model) is important in order to clarify
the viability of this class of
models.
Clearly, the allowed range of $q$ depends on the mass of the
scalar field and therefore on the parameter $\beta$ as well as on
the introduction of the parameter $\omega$ which could stabilise
the scalar field through the kinetic term (see eq.
(\ref{bdbsbm})). Constraints on
the dilatonic coupling in extensions of the standard electroweak
model \cite{Kimberly:2003rz,Shaw:2004hk} have recently been imposed by the LHC \cite{Coleppa:2011zx,Lahanas:2012hd,Bellazzini:2012vz}.
\end{itemize}
In conclusion, the existence of a dilatonic
coupling in field theories predicting the existence of topological
defects implies the presence of interesting new properties for the
predicted defects which makes these models worth of further
investigation.

{\bf Acknowledgements}

This research has been co-financed by the European Union (European
Social Fund - ESF) and Greek national funds through the Operational
Program ``Education and Lifelong Learning'' of the National Strategic
Reference Framework (NSRF) - Research Funding Program: ARISTEIA.\@
Investing in the society of knowledge through the European Social

\end{document}